# Transient Non-linear Thermal FEM Simulation of Smart Power Switches and Verification by Measurements

Vladimír Košel[1,2], Roland Sleik[1], Michael Glavanovics[1]

[1] KAI Kompetenzzentrum Automobil- und Industrieelektronik GmbH
Europastraße 8, Villach, 9524 Austria
[2] Slovak University of Technology in Bratislava,
Faculty of Electrical Engineering and Information Technology

*Abstract*- **Thermal FEM (Finite Element Method) simulations can be used to predict the thermal behavior of power semiconductors in application. Most power semiconductors are made of silicon. Silicon thermal material properties are significantly temperature dependent. In this paper, validity of a common non-linear silicon material model is verified by transient non-linear thermal FEM simulations of Smart Power Switches and measurements. For verification, over-temperature protection behavior of Smart Power Switches is employed. This protection turns off the switch at a pre-defined temperature which is used as a temperature reference in the investigation. Power dissipation generated during a thermal overload event of two Smart Power devices is measured and used as an input stimulus to transient thermal FEM simulations. The duration time of the event together with the temperature reference is confronted with simulation results and thus the validity of the silicon model is proved. In addition, the impact of non-linear thermal properties of silicon on the thermal impedance of power semiconductors is shown.**

## I. Introduction

In automotive applications, Smart Power semiconductors are exposed to extraordinary thermal stress with average junction temperature reaching 150°C in normal operation and well beyond 200°C during fault conditions [1]. In addition, high speed periodic switching of inductive loads, such as fuel injector valves, causes additional power dissipation pulses, driving the peak junction temperature up to 200°C for millisecond time intervals, affecting the device's reliability over its life time [2]–[5]. Accurate evaluation of electrical and thermal stresses is thus crucial to ensure reliable power switch operation within the specified safe operating area, given by maximum permissible voltage, current and junction temperature.

The pressure of cost reduction leads to shrinkage of chip area. In order to make smaller devices possible, new technologies that significantly reduce chip area while preserving the same electrical resistance and nominal current of the power switch are developed. The consequences are an increase of power density and higher thermal resistance of power devices. Taking these two factors into consideration, today's power semiconductors are exposed to thermal stresses significantly higher than their predecessors. This can result in reduced lifetime and in extreme cases, thermal destruction of power devices.

In order to create robust design for power semiconductor devices, it is important to consider not only electrical, but also thermal aspects. This can be done during the product design phase by applying thermal FEM simulations. The computing power is nowadays sufficient to perform very accurate and complex FEM simulations within reasonable time.

Main inputs to a FEM simulator are geometrical data and material properties. Most materials used for production of power semiconductors change their thermal properties with temperature. The hotspot of a device is usually located in the silicon die. Others structures surrounding the die have much lower temperature. Therefore for many materials their temperature dependence can be neglected. For silicon the situation is much worse, first of all because the die is in a region with highest temperate and secondly because thermal material properties of silicon are significantly temperature dependent. Unfortunately the temperature dependence is even going in an undesirable direction. Silicon thermal resistance and specific heat capacity increase with temperature. This causes a positive thermal feedback. The higher the temperature the higher the resistance and vice versa. Therefore the thermal material properties of silicon have to be carefully taken into account. Previous work has

This work was jointly funded by the Federal Ministry of Economics and Labour of the Republic of Austria (contract 98.362/0112-C1/10/2005) and the Carinthian Economic Promotion Fund (KWF) (contract 98.362/0112-C1/10/2005).

Corresponding author: vladimir.kosel@k-ai.at, Vladimir Košel 2003 graduated in electrical engineering at Slovak University of Technology (STU) in Bratislava, Faculty of Electrical Engineering and Information Technology (FEI). He started his PhD in 2003 at STU FEI, and worked on reliability of Smart Power Switches at Infineon Technologies in Villach, Austria. Since 2006, he is continuing his PhD project at Kompetenzzentrum Automobil- und Industrie- Elektronik in Villach.

adressed this issue based on nonlinear, temperature dependent dynamic compact (i.e. RC-chain based) models along with measurements of thermal impedance [11].

In this paper the relevance of a non-linear silicon material model for transient thermal FEM simulations of power semiconductors is proved by an indirect method using both thermal FEM simulations and measurements. In addition the influence of the silicon thermal properties on thermal impedance of Smart Power Switches (SPS) is shown.

## II. SIMULATION PRE-PROCESSING

### A. Modeling of silicon thermal properties

The thermal properties of silicon are significantly dependent on temperature. Many investigations have been done on this subject [7]–[9]. The results vary depending on dimensions and manufacturing process of the silicon sample as well as on the measurement method. Comparing results of [6], [8] and [9], in general, one may expect that thin silicon layers have a different thermal conductivity than bulk crystalline silicon. Whether the thermal material model of bulk silicon is valid for power semiconductors with a die thickness of about 200 µm is investigated in this paper.

The bulk silicon thermal material properties can be described piecewise by empirical functions based on measurements presented in [10] and having e.g. the following form [6]:

$$k(T) = \begin{cases} 2.025 \cdot 10^6 \cdot T^{-1.675} & T < 273K \\ 2.475 \cdot 10^5 \cdot T^{-1.3} & T > 273K \end{cases} \quad (1)$$

$$c_p(T) = \begin{cases} 2.798 \cdot T - 27.96 & T < 273K \\ 2.428 \cdot 10^2 \cdot T^{0.195} & T > 273K \end{cases} \quad (2)$$

where $T$ is the temperature in K, $k(T)$ in W/m·K and $c_p(T)$ in J/kg·K represent the temperature dependent thermal conductance and specific heat capacity, respectively.

### B. FEM modeling of Smart Power Switches

For our study, the FEM simulator FlexPDE is used [13]. Two Smart Power reference devices are modeled. Their three dimensional (3D) models are depicted in Fig. 1 and 2. In general a SPS product consists of one or more dies carrying one or several power switches covered with power metallization, die attach, heatsink (power package) or leadframe (plastic package) along with pins, bonding wires; all these elements are encapsulated in Epoxy molding compound.

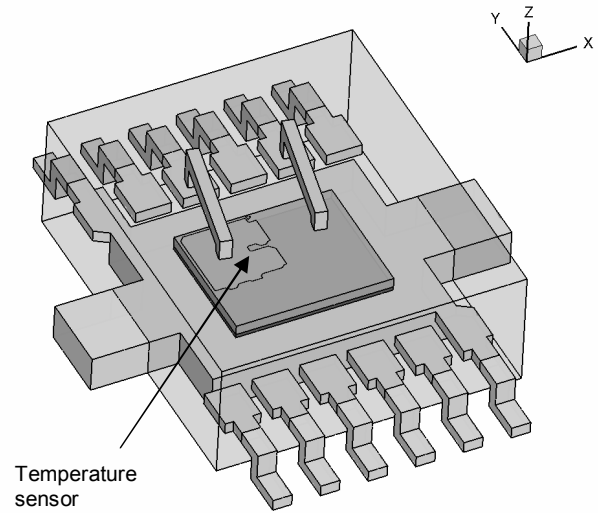

Temperature sensor

Fig. 1. FEM model of reference device #1

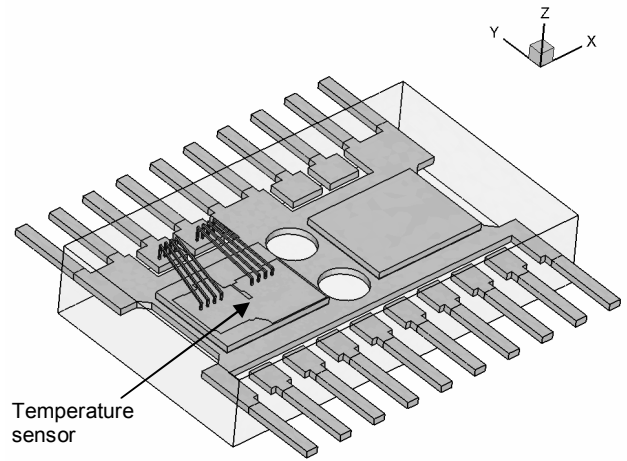

Temperature sensor

Fig. 2. FEM model of reference device #2

TABLE I
MODELLDED PARTS OF BOTH DEVICES AND ASSIGNED MATTERIALS

| Device | Part | Material |
|--------|------|----------|
| #1, #2 | Die & epitaxial layer | Silicon |
| #1, #2 | Power metallization | Aluminum |
| #2 | Die attach - glue | CRM 1033B |
| #1 | Die attach - solder | PbSn2Ag2.5 |
| #2 | Leadframe | Copper |
| #1 | Heatsink | Copper |
| #1 | Molding compound | KMC 289 |
| #2 | Molding compound | KMC 165-8 |

The thermal properties of molding compound and die attach of both devices were obtained from the material manufacturers' specifications. For heatsink, leadframe, bonding wires and power metallization, standard properties of copper and aluminum were used. Table I lists considered physical elements of both devices in our simulation models and materials corresponding to them. The silicon thermal material properties are described by equations (1) and (2).

To save modeling and computation time, the geometry of simulated devices is adequately simplified. The package outlines are modeled as cuboids; the pins and power bonding wires as prismatic structures. The signal bonding wires have no physical contact with any active area of power transistors and therefore do not influence directly their temperature. This allows us to omit them in the models.

Great emphasis is put on the modeling of chip and power transistor geometry. In our models, the chips consist of several layers: silicon substrate, epitaxial layer, power metallization.

A power switch usually consists of many MOSFET cells which can be regarded as a large array of very small transistors connected together in parallel [12]. The area of one cell in our sample technology is approximately 25 μm² which is small enough in comparison to the whole power transistor of about 2 mm². Modeling exactly the inside of every cell, a huge number of elements would be generated and the problem would not be solvable in reasonable time. To avoid this, it is assumed that the power dissipation density in all cells has the same value. This allows us to consider the power transistor area as a homogenuously thermally active region. The power dissipation density which is an input parameter to the simulator is then calculated as a ration of the total power dissipation to the active layer volume in the given power transistor. The active volume is calculated by multiplication of the active area with the thickness of the epitaxial layer. The shape of the active area of a power transistor is extracted from the chip layout. These shapes form the regions containing thermal sources in the FEM model and are set into the epitaxial layer immediately underneath the power metallization.

## III. NON-LINEAR THERMAL FEM SIMULATIONS

To verify the validity of the non-linear silicon model in the temperature range from 20°C to 200°C, the following approach was used. The chosen SPS reference devices incorporate a broad range of smart functions, including an over-temperature protection together with a temperature sensor based on a *pn* junction. The sensor is placed in a cutout located in the active area of the power transistor. In overload conditions, e.g. if a high current flows through the switch, the dissipated power in the switch may cause a sudden and excessive temperature rise. If the temperature at the sensor exceeds a predefined temperature shutdown threshold, the thermal protection responds by turning the switch off.

Let us assume a real overload event of an SPS leading to the thermal shut down. During this event the dissipated power in the switch causes a temperature rise inside the package. In certain time the temperature on the temperature sensor reaches the temperature shutdown threshold and the thermal protection turns the switch off. - An overload event leading to thermal shutdown is a suitable scenario to verify the validity of the silicon material model. The following hypothesis is proposed: if the observed event is simulated based on measured shutdown time and power dissipation and if the proposed FEM model is valid, then the calculated sensor temperature at time of thermal shut down has to be equal to the measured temperature shutdown threshold.

### A. Measurements on Smart Power Switches

As device suppliers usually only specify a minimum and maximum value of the temperature shutdown threshold $T_{sdth}$, several measurements on the reference SPS devices were performed to determine its accurate value. The following explains the measurement methodology. The SPS device is driven in a regular operating condition with a small load current through the investigated channel to keep the dissipated power as small as possible. This avoids any significant local temperature rise on the temperature sensor and thus helps to minimize the measurement error. Next, the case temperature of the device is increased very slowly till the thermal shutdown occurs. This measurement procedure was repeated on several samples of both reference devices to verify reproducibility. The averaged results are listed in Table II.

TABLE II
MEASURED TEMPERATURE SHUTDOWN THRESHOLD

| Product | Device #1 | Device #2 |
|---------|-----------|-----------|
| $T_{sdth}$ [°C] | 170.5 | 172.2 |

Thermal shut down can also be activated dynamically through a short circuit condition. Both reference SPS devices were switched on with a very low impedance load. The high current flowing through the switch and the corresponding voltage drop over the power transistor were recorded until thermal shutdown occured. Power dissipation waveforms were then calculated from these time records (Fig. 3 and 4).

### B. FEM simulations of thermal shutdown event

The goal of the FEM simulations is to determine the temperature on the temperature sensor under the short circuit conditions mentioned above. The measured waveforms of power dissipation are used as inputs to FEM simulations, for each device respectively

The sensors have a rectangular shape with a length of about 84 μm and 250 μm for reference devices #1 and #2, respectively. The width of both sensors is much smaller than their length and therefore the sensors are considered as a line structure in the FEM model. However, the lengths of these sensors are not negligible compared to the dimensions of the power transistors on the die. Therefore certain temperature differences across the temperature sensor may be expected.

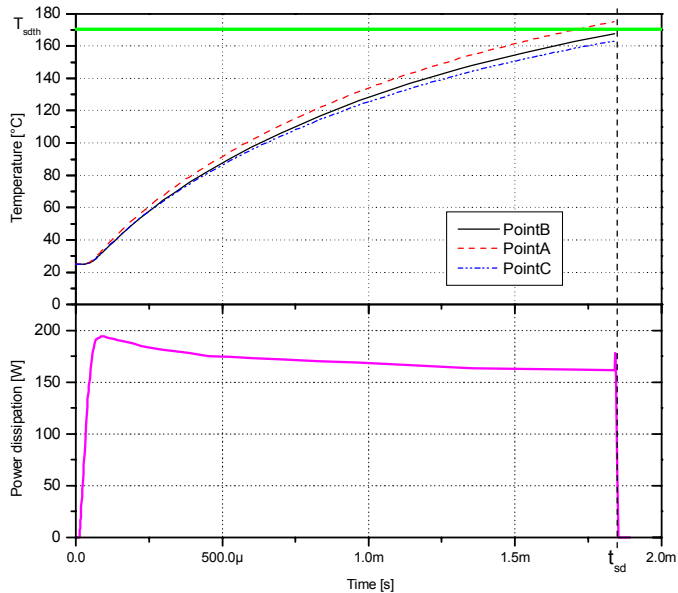

Fig. 3. Device #1 – Measurements and FEM simulation of overload event;
*upper* - measured shutdown temperature threshold and calculated temperature
over the sensor observed in three points in the simulation;
*lower* – measured power dissipation in the switch

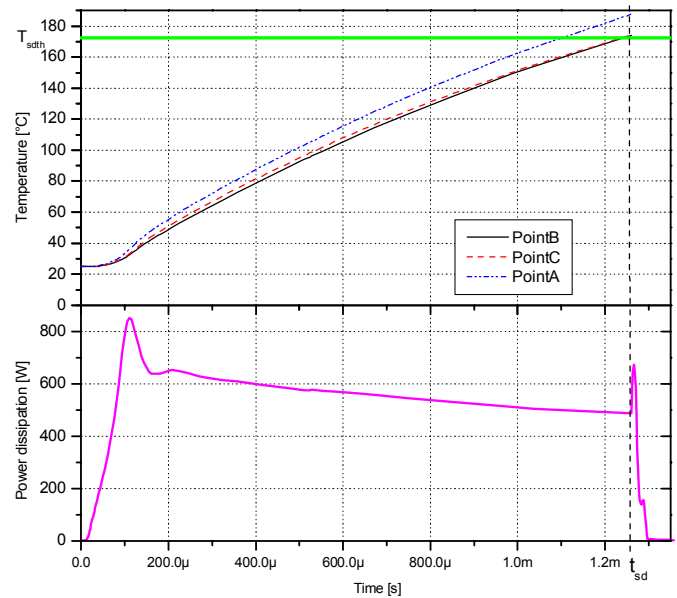

Fig. 4. Device #2 – Measurements and FEM simulation of overload event;
*upper* - measured shutdown temperature threshold and calculated temperature
over the sensor observed in three points in the simulation;
*lower* – measured power dissipation in the switch

To determine the mean sensor temperature, three points, two at the sensor ends (points A, C) and one in the middle (point B) have been observed in the simulations. The point A is closest to the hot spot at the center of the power transistor.

The simulation results of reference devices #1 and #2 are plotted in Fig. 3 and 4, respectively. The non-uniform temperature distribution over the sensor can be clearly seen. The mean temperature of the sensor has been calculated by discrete integration with Kepler's rule. The calculated sensor temperatures at shutdown time are listed in Table III. The percentage deviations also listed in Table III are calculated between the mean values derived from FEM simulation and the measured temperature shutdown thresholds for every device respectively. The deviation between simulation and measurement is below 3% for both devices, which confirms the validity of the non-linear silicon model for the investigated temperature range.

TABLE III
SIMULATION RESULTS

| Product | Point | A | B | C | Average | Deviation % |
|---------|-------|---|---|---|---------|-------------|
| Device #1 | $T$ [°C] | 175.2 | 167.6 | 163.2 | 168.1 | -1.3 |
| Device #2 | | 174.4 | 174.6 | 188.0 | 176.7 | +2.6 |

## IV. INFLUENCE OF SILICON NON-LINEAR THERMAL PROPERTIES ON THERMAL IMPEDANCE

Some thermal properties of power semiconductor products are usually specified in their datasheets. This information helps the electronic system designers to calculate the temperature of a device in their application. The most commonly specified parameter is the thermal resistance between the junction of a power transistor (switch) and its case, usually denoted as $R_{thJ-C}$. If operating conditions are known, the junction temperature rise $\Delta T_j$ in the device can be calculated from the mean dissipated power $P_{mean}$ with following formula

$$\Delta T_j = P_{mean} \cdot R_{thJ-C} . \qquad (3)$$

For general dynamic cases the so-called thermal impedance (i.e. the step response of the junction temperature when 1W of constant power is applied) is more interesting. Knowing this parameter, the temperature rise over time can be calculated for arbitrary power dissipation, e.g. a junction temperature rise of power transistor switching an incandescent lamp. For a constant power pulse, junction temperature rise is expressed as follows

$$\Delta T_j(t) = P_{pulse} \cdot Z_{th}(t) . \qquad (4)$$

The reference device #1 was chosen to analyze the influence of the silicon non-linear thermal properties on thermal impedance. Transient non-linear thermal simulations at ambient temperature for different dissipated power were performed. Output of these simulations is the development of junction temperature distribution over time.

TABLE IV
THERMAL RESISTANCE

| Power [W] | 1 | 50 | 100 |
|---|---|---|---|
| $R_{th,J-C}$ [K/W] | 1.67 | 1.87 | 2.11 |

The thermal impedance for every simulated case was calculated by (4), where $\Delta T_j$ is here understood as the peak junction temperature. The results are plotted in Fig. 5. The maximum value of thermal impedance for large time corresponds to the static thermal resistance. Thermal resistances for different dissipated powers are listed in Table IV. The results show a drastic change of the thermal impedance depending upon the power dissipation. Approximately after 200 µs all three curves start to diverge. In the investigated power dissipation range the thermal resistance changes from 1.67 K/W to 2.11 K/W, which is equivalent to a 26 % increase due to the non-linear thermal properties of silicon.

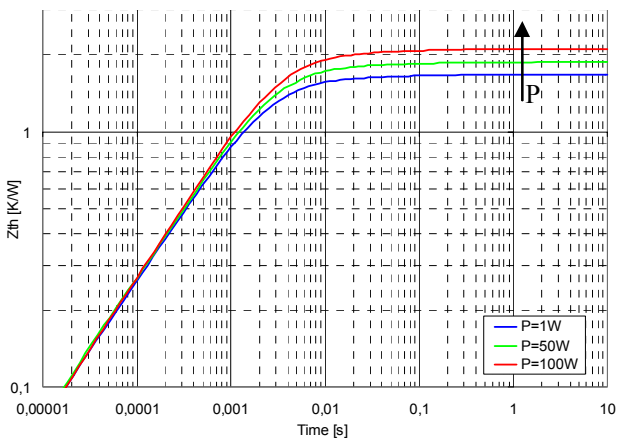

Fig. 5. Thermal impedance of device #1 for different power dissipation

## IV. CONCLUSION

The validity of FEM models of Smart Power Switches based on non-linear silicon thermal properties was proved by measurements on two SPS reference devices along with corresponding FEM simulations. In this investigation the over-temperature protection function of SPS devices was used. The temperature shutdown threshold of the thermal overload protection was measured for both reference devices and used as a temperature reference. A short circuit event in the SPS devices leading to thermal shutdown was then chosen as a suitable scenario for FEM model verification. The waveforms of dissipated power under short circuit conditions were acquired by measurements and used as a stimulus to the transient non-linear thermal FEM simulations for both devices. The simulated sensor temperatures at thermal shutdown show a very good match with static measurements of the temperature shutdown threshold and thus confirm the validity of the non-linear thermal FEM model over a temperature range from 20°C to 200°C.

The influence of the non-linear silicon properties on thermal impedances for different dissipated powers is demonstrated on one reference device, again using non-linear thermal FEM simulations. The results show a significant increase of thermal impedance with power dissipation. To perform realistic thermal FEM simulations of power semiconductors, the non-linear thermal properties of silicon have thus to be taken into account.

### ACKNOWLEDGMENT

The authors would like to thank H. Grünbacher from KAI GmbH. Our thanks go also to U. Fröhler, F. Riedl, B. Wang, D. Härle and Ch. Schreiber from Infineon Technologies and all others who have contributed to this work.